\documentstyle[preprint,eqsecnum,aps,tighten,epsf]{revtex}

\begin{document}

\newcommand{\dfrac}[2]{\frac{\displaystyle #1}{\displaystyle #2}}

%%%%%%%%%%%%%%%%%%%%%%%%%%%%%%%%%%%%%%%%%%%%%%%%%%%%%%%%%%%%%%%%%%%%%%%%%%%%%
\draft
\preprint{VPI--IPPAP--00--01}

\title{Constraints on Two--Higgs Doublet Models at Large 
${\tan{\beta}}$ from W and Z decays}
\author{
Oleg~Lebedev\thanks{electronic address: lebedev@quasar.phys.vt.edu},
Will~Loinaz\thanks{electronic address: loinaz@alumni.princeton.edu}, and
Tatsu~Takeuchi\thanks{electronic address: takeuchi@vt.edu}
}
\address{Institute for Particle Physics and Astrophysics,
Physics Department, Virginia Tech, Blacksburg, VA 24061}

\date{February 9, 2000}
\maketitle

\begin{abstract}
We study constraints on type--II two Higgs doublet models at large $\tan\beta$
from LEP/SLD $Z$--pole data and from lepton universality violation in 
$W$ decay.
We perform a global fit and find that, in the context of $Z$ decay,
the LEP/SLD experimental values for lepton universality violation, 
$R_b$, and $A_b$ all somewhat disfavor the model.
Contributions from the neutral Higgs sector can be used to
constrain the scalar--pseudoscalar Higgs mass splittings.  
Contributions from the charged
Higgs sector allow us to constrain the charged Higgs mass.   
For $\tan\beta=100$ we obtain the $1\sigma$ classical (Bayesian) bounds of
\[
m_{H^\pm}\geq 670\,{\rm GeV} \; (370\,{\rm GeV})
\quad\mbox{and}\quad
1 \geq m_{h^0}/m_{A^0} \geq 0.68 \; (0.64).
\]
The $2\sigma$ bounds are weak.
Currently, the Tevatron experimental limits on lepton universality violation 
in $W$ decay provide no significant constraint on the Higgs sector.
\end{abstract}


%%%%%%%%%%%%%%%%%%%%%%%%%%%%%%%%%%%%%%%%%%%%%%%%%%%%%%%%%%%%%%%%%%%%%%%%%%%%%%
\narrowtext

\section{Introduction}

Perhaps the most important unanswered question in particle physics today is:
``What is the nature of electroweak symmetry breaking?''  
The Standard Model (SM) incorporates the simplest mechanism:
a Higgs sector consisting of a
single self--interacting scalar $SU(2)$ doublet of hypercharge $Y=1$.
Upon breaking of electroweak symmetry, the physical spectrum of the 
SM Higgs sector consists of one CP--even neutral Higgs particle.  
Current experimental data do not definitively contradict the SM, 
but persistent deviations in precision 
electroweak data from SM predictions on the edge of statistical significance 
tantalize with the possibility of new physics.  This, together
with various theoretical prejudices which suggest that the SM cannot be 
a complete theory, motivates the detailed study of alternative scenarios 
of eletroweak symmetry breaking (EWSB).

The Two Higgs Doublet Model (2HDM) \cite{Haber:1997dt}
is the most straightforward extension of the EWSB mechanism of the SM.
The theory proposes a pair of scalar
$SU(2)$ doublets, both with hypercharge $Y=1$.
Depending on the version of the 2HDM, these scalars may couple in 
various ways to the quarks and leptons.  After electroweak 
symmetry is broken, the spectrum of the Higgs sector consists of five physical
Higgs bosons: two neutral CP--even scalars ($h_0$ and $H_0$), 
a neutral CP--odd scalar ($A_0$), and a pair of charged scalars ($H^\pm$).
These particles could be detected via direct production at colliders, 
but their effects may also be visible indirectly, 
through their contributions as intermediate states in decay processes.

In this paper we consider the indirect signatures of the 2HDM in 
flavor--conserving W and Z-decays through its contribution to decay 
amplitudes via loop corrections.  
We consider only Type--II 2HDM models, 
in which the $I_3=\frac{1}{2}$ fermions couple to one Higgs doublet 
and the $I_3=-\frac{1}{2}$ fermions couple to the other.  
We also focus on the large $\tan\beta$ region{\footnote{Perturbativity of the
$b$ and $t$ Yukawa couplings requires $0.3 \leq \tan\beta \leq 120$ 
(see, for example \cite{Haber:1997dt}). }, 
in which the Higgs couplings to the down--type quarks and the 
charged leptons are enhanced{\footnote{
This model is often studied embedded in the Minimal Supersymmetric extension 
of the Standard Model (MSSM) \cite{Loinaz:1998ph}, 
although we do not consider it in this context 
here.}}.
This can potentially lead to
observable (or constrainable) flavor dependent corrections in 
$Z$ and $W$ decay, especially for the third generation ($b$ and $\tau$).

One--loop corrections to flavor--conserving $Z$ decays in the 2HDM
have been considered previously in Refs.~\cite{Denner:1991ie,Grant:1995ak,%
Hisano:1997fx,Logan:1999if,Haber:1999zh},
including as a possible explanation for the now--defunct `$R_b$ anomaly'.  
The $Z$--pole runs at LEP and SLD are complete and essentially 
all of the data have been analyzed.  
The `$R_b$ anomaly' has disappeared only to be replaced by the 
`$A_b$ anomaly' \cite{Chanowitz:1999jj,Field:1999ng};
thus, it is timely to revisit the model.  
We perform a global fit to all LEP/SLD $Z$--pole observables, 
and we examine the competing constraints from lepton universality, 
$R_b$, and $A_b$ on the charged and neutral sectors of the model.  
In addition, we study constraints on the model from 
lepton universality violation in $W$ decays, 
which have not been previously considered.

%%%%%%%%%%%%%%%%%%%%%%%%%%%%%%%%%%%%%%%%%%%%%%%%%%%%%%%%%%%%%%%%%%%%%%%%%%%%%%

\section{Leptonic W decays}

In this section we calculate the constraints on the large--$\tan\beta$ 2HDM 
from lepton universality violation in $W$ decays.
We use the Feynman rules and conventions of Ref.~\cite{Gunion:1989we}.  
Our notation for the scalar and tensor integrals is  
established in Ref.~\cite{Lebedev:1999vc}.

The leading (in $\tan\beta$) one--loop corrections to the
decay $W^-\rightarrow \tau^- \bar{\nu}_\tau$ are shown in 
Fig.~\ref{Wdiagrams}.
The corresponding contributions to the amplitude are
\footnote{In computing the one--loop vertex corrections in both $W$ and $Z$
decays, it is a good approximation to neglect light fermion masses in loops.}:
\begin{eqnarray}
\lefteqn{
-\frac{g^2}{4}\left( \frac{m_\tau \tan\beta}{m_W} \right)^2
\left[ -i\frac{g}{\sqrt{2}}
       W^\mu(Q)\,\bar{\tau}(p)\gamma P_L \nu_\tau(q)
\right] \times} \qquad\qquad & & \cr
(1a/h^0) 
& : & \sin^2\!\alpha\; 
      2\hat{C}_{24}(0,0,Q^2;0,m_{h^0},m_{H^\pm}) \cr
(1a/H^0)
& : & \cos^2\!\alpha\;
      2\hat{C}_{24}(0,0,Q^2;0,m_{H^0},m_{H^\pm}) \cr
(1a/A^0)
& : & 2\hat{C}_{24}(0,0,Q^2;0,m_{A^0},m_{H^\pm}) \cr
(1b/h^0)
& : & \sin^2\!\alpha\;\;B_1(0;0,m_{h^0}) \cr
(1b/H^0)
& : & \cos^2\!\alpha\;\;B_1(0;0,m_{H^0}) \cr
(1b/A^0)
& : & B_1(0;0,m_{A^0}) \cr
(1c)
& : & 2B_1(0;0,m_{H^\pm})
\end{eqnarray}
with $Q^2=m_W^2$.
The tree level amplitude is the expression in the square brackets.
For the diagrams involving $h^0$ and $H^0$, we have dropped terms
subleading in $\tan\beta$\footnote{The 
subleading contributions of the $h^0$ and $H^0$ diagrams
combine with the diagrams involving the Goldstone bosons to give finite
results.}.  In the above we have made the
large $\tan\beta$ approximations:
\begin{eqnarray}
\cos(\beta - \alpha) & \approx & \sin{\alpha} \cr
\sin(\beta - \alpha) & \approx & \cos{\alpha} \cr
\sin\beta & \approx & 1 \cr
\cos\beta & \approx & 0.
\label{eq:trigapprox}
\end{eqnarray}

Combining the above corrections, with factors of $1/2$ for the wave--function
renormalization diagrams (1b) and (1c), leads to a shift in the 
$W\tau\bar{\nu}_\tau$ coupling given by:
\begin{eqnarray}
\frac{ \delta g_\tau }{ g }
& = & -\frac{g^2}{2}
       \left( \frac{ m_\tau \tan\beta }{ m_W }
       \right)^2 \times  \cr
&   &
\Biggl[\;\sin^2\!\alpha
       \left\{ \hat{C}_{24}(0,m_{h^0},m_{H^\pm})
               + \frac{1}{4}\,B_1(0,m_{h^0})
               + \frac{1}{4}\,B_1(0,m_{H^\pm})
       \right\}
\Biggr.  \cr
&   & \quad
     + \cos^2\!\alpha
       \left\{ \hat{C}_{24}(0,m_{H^0},m_{H^\pm})
               + \frac{1}{4}\,B_1(0,m_{H^0})
               + \frac{1}{4}\,B_1(0,m_{H^\pm})
       \right\} \cr
&   & \quad
\Biggl.
     + \left\{ \hat{C}_{24}(0,m_{A^0},m_{H^\pm})
               + \frac{1}{4}\,B_1(0,m_{A^0})
               + \frac{1}{4}\,B_1(0,m_{H^\pm})
       \right\} 
\Biggr]
\label{eq:Wcouplingshift}
\end{eqnarray}
where we have suppressed the external momentum dependence of the 
integrals for notational simplicity.
Similar shifts to the $W\mu\bar{\nu}_\mu$ and $We\bar{\nu}_e$ vertices 
exist but they are suppressed by factors of $(m_\mu/m_\tau)^2$ and
$(m_e/m_\tau)^2$ so we neglect them.

The complete expression for the finite combination of integrals
seen in the curly brackets of Eq.~(\ref{eq:Wcouplingshift}), namely,
\begin{equation}
\zeta(Q^2;m_1,m_2) \equiv
\hat{C}_{24}(0,0,Q^2;0,m_1,m_2) 
+ \frac{1}{4}\,B_1(0;0,m_1) + \frac{1}{4}\,B_1(0;0,m_2),
\label{eq:zetadef}
\end{equation}
can be found in the appendix of Ref.~\cite{Lebedev:1999vc}.  
However, for our purposes it will suffice to expand it
in powers of $Q^2=m_W^2$:
\begin{eqnarray}
\zeta(m_W^2;m_1,m_2)
& = &
-\frac{1}{(4\pi)^2} \frac{1}{4}\,
G\!\left( \frac{m_1^2}{m_2^2} \right) \cr
&   &
-\frac{1}{(4\pi)^2} 
\frac{m_W^2}{12(m_1^2-m_2^2)^2}
\Biggl[ m_1^2 + m_2^2 - \frac{2 m_1^2 m_2^2}{m_1^2 - m_2^2} 
                        \ln \frac{m_1^2}{m_2^2} 
\Biggr] \cr
&   & + \cdots
\label{eq:expn}
\end{eqnarray}
where 
\begin{equation}
G(x) \equiv 1 + \frac{1}{2} \left( \frac{1+x}{1-x} \right) \ln{x}.
\label{eq:Gdef}
\end{equation}
Observe that the function $G(x)$ is negative semi--definite so that
the leading term is non--negative for all masses $m_1$ and $m_2$.
This term dominates the subleading term unless the splitting between
$m_1$ and $m_2$ is less than about $m_W/2$.
In the limit that the masses are degenerate, 
the leading term vanishes and the expansion reduces to
\begin{equation}
\zeta(m_W^2;m,m)
= -\frac{1}{(4\pi)^2} \frac{m_W^2}{36 m^2} + \cdots
\end{equation}
In the limit $m_1\rightarrow \infty$ 
(the full expression is symmetric in $m_1$ and $m_2$)
the expression becomes
\begin{equation}
\zeta(m_W^2;m_1,m_2)
\Longrightarrow
-\frac{1}{(4\pi)^2} \frac{1}{4} \biggl(
1 + \frac{1}{2} \ln{\frac{m_2^2}{m_1^2}}
\biggr) + \cdots
\end{equation}
Thus, Eq.~(\ref{eq:Wcouplingshift}) appears to lead to non--decoupling of 
heavy particles. 
That is, when $m_{A^0}$ and $m_{H^\pm}$ are taken to be large independently, 
the amplitude does not vanish. 
However, in the general 2HDM the mass eigenvalues and mixing angle are related 
in such a way that if $ m_{A^0}\rightarrow \infty$ while the couplings and 
the $W$ mass are held fixed, 
the Higgs masses and mixing angle approach the limit \cite{Haber:1997dt}:
\begin{equation}
m_{H^0} \simeq  m_{H^\pm} \simeq m_{A^0} \quad ; \quad 
m_{h^0}/m_{A^0} \rightarrow 0 \quad ; \quad 
\cos(\alpha-\beta) \rightarrow 0.
\end{equation}
In this limit the amplitude vanishes and decoupling is obtained.
This decoupling behavior can be understood as follows:  
In the large $\tan\beta$ limit, $\sin\alpha \rightarrow 0$ and the 
two Higgs doublets do not mix. 
Since large $\tan\beta$ implies $v_1 \rightarrow 0$, 
electroweak symmetry is unbroken in the $\Phi_1$ sector.  
The leading $\tan^2\beta$ diagrams are then due to the 
$W^- \Phi_1^{-*} \Phi_1^0$ vertex and corresponding 
wave function renormalization diagrams ({\it i.e.} with
$\Phi_1^0$ and $\Phi_1^-$ in the loop).  
The sum of these diagrams vanishes at $p^2=0$ as a consequence of the 
Ward identity, and thus the heavy Higgs bosons
decouple.\footnote{The leading $\tan\beta$ contribution of the $h^0$ boson 
does not exhibit decoupling by itself : it is proportional to 
$\tan^2\!\beta\; \sin^2\!\alpha \; \ln m_{A^0}^2 \rightarrow \ln m_{A^0}^2$ 
since $\sin\alpha \sim -\cos\beta + {\cal{O}}(m_Z^2/{m_{A^0}^2})$
in the decoupling limit \cite{Haber:1997dt}. As the result is independent of
$\tan\beta$, subleading $\tan\beta$ diagrams must be included to obtain
the decoupling behavior.}

Setting $m_{A^0} = m_{H^0} = m_{H^\pm} \equiv m$ and 
neglecting the $m_{h^0}$ contribution,
the shift in the coupling, Eq.~(\ref{eq:Wcouplingshift}), 
is small and positive:
\begin{eqnarray}
\frac{\delta g_\tau}{g}
& = &
-\frac{g^2}{2} \left( \frac{m_\tau \tan\beta}{m_W} \right)^2 
\times 2\zeta(m_W^2;m,m) \cr
& \approx & 
-\frac{g^2}{2} \left( \frac{m_\tau \tan\beta}{m_W} \right)^2 
\left\{ -\frac{1}{(4\pi)^2} \frac{m_W^2}{18 m^2} 
\right\} \cr
& = & \left( \frac{g\, m_\tau \tan\beta}{24 \pi m} \right)^2.
\label{eq:degenerateW}
\end{eqnarray}
Away from this limit, the shift in the coupling is negative.
\begin{eqnarray}
\frac{\delta g_\tau}{g} 
& = &
\frac{1}{(4\pi)^2}\frac{g^2}{8}
\left( \frac{m_\tau \tan\beta}{m_W} \right)^2 \cr
&   & \times
\left[\;
  \sin^2\!\alpha\;\;G\!\left( \frac{m_{H^\pm}^2}{m_{h^0}^2} \right) 
+ \cos^2\!\alpha\;\;G\!\left( \frac{m_{H^\pm}^2}{m_{H^0}^2} \right) 
+ G\!\left( \frac{m_{H^\pm}^2}{m_{A^0}^2} \right)\;
\right] \cr
& \leq & 0. \phantom{\frac{1}{2}}
\end{eqnarray}
So, the model predicts a negative $\delta g_\tau$, except in the 
limit that the Higgs mass splittings are small ($\leq m_W/2$).
The magnitude of the shift is maximal for an extreme non--decoupling 
case in which the charged Higgs is much heavier than the neutral Higgses.  
In this case it reduces to
\begin{equation}
\frac{\delta g_\tau}{g} =
\frac{1}{(4\pi)^2} \frac{g^2}{4} 
\left( \frac{m_\tau \tan\beta}{m_W}
\right)^2\;
G\!\left( \frac{ m_{H^\pm}^2 }{ m_0^2 } 
      \right)
\label{eq:simpleWshift}
\end{equation}
if we assign a common mass, $m_0$, to the neutral Higgses.

The current bound on lepton universality violation in leptonic 
$W$ decays from the D{0\kern-6pt/} Collaboration is \cite{Rimondi:1999pu}
\begin{eqnarray}
\frac{ g_\tau }{ g_e } = 1.004 \pm 0.019\,(stat.) \pm 0.026\,(syst.).
\label{eq:Wdata}
\end{eqnarray}
The central value of $\delta g_\tau$ is positive, which is not allowed
when the leading $G(x)$ term dominates.
However, this fact is inconclusive since the experimental error is large.
Using $\overline{m}_\tau(m_W) = 1.777\;{\rm GeV}$ and 
$2m_W/g = v = 246\,{\rm GeV}$, 
and adding systematic and statistical errors in quadrature, we obtain 
from Eqs.~(\ref{eq:simpleWshift}) and (\ref{eq:Wdata}):
\begin{equation}
G\!\left(\frac{m_{H^\pm}^2}{m_0^2} \right) = 
\left( \frac{100}{\tan\beta} \right)^2
\Bigl[\; 1.2 \pm 9.7 \;\Bigr] 
\end{equation}
Since $G(x)$ negative semi--definite and invariant under 
$x \leftrightarrow 1/x$, at $\tan\beta=100$ this leads to a 1$\sigma$ bound of
\[
G\!\left( \frac{m_{H^\pm}^2}{m_0^2} \right) > -8.5
\]
which translates to
\[
\frac{m_0}{m_{H^\pm}}\quad\mbox{or}\quad\frac{m_{H^\pm}}{m_0} 
< 1.3 \times 10^4.
\]
For smaller $\tan\beta$ the bound is even weaker.  

Similarly, if we assume the limit of Eq.~(\ref{eq:degenerateW}), 
the best--fit value of the common mass is (in GeV)
\begin{equation}
\left( \frac{ 100\,{\rm GeV} }{ m } \right)^2 = 
\left( \frac{ 100 }{ \tan\beta } \right)^2  
\Bigl[\; 17 \pm 132 \;\Bigr]
\end{equation}
At 1$\sigma$ and $\tan\beta=100$, this translates into
\[
m > 8\,{\rm GeV}
\]
so, the bound is extremely weak in this mass-degenerate limit as well.
Thus, even for $\tan\beta = 100$ the current data gives 
no significant $1\sigma$ constraint on the Higgs masses.

%%%%%%%%%%%%%%%%%%%%%%%%%%%%%%%%%%%%%%%%%%%%%%%%%%%%%%%%%%%%%%%%%%%%%%%%%%%%%%

\section{Constraints from LEP/SLD Observables}

In this section we perform a global analysis of LEP/SLD precision 
electroweak data in the context of the large $\tan\beta$ 2HDM.  
We calculate the linearized shifts in the $Z f \bar{f}$ couplings
from SM predictions, fit these shifts to the data, and use the 
results of the fit to constrain model parameters.

\subsection{Corrections to the couplings}

As in the $W$ decay case, large $\tan\beta$ enhances the coupling
of the Higgs sector to charged leptons and down--type quarks,
but even then one only needs to consider the third generation fermions.
Below we list corrections to $Z\rightarrow b\bar{b}$, $\tau\bar{\tau}$,
$\nu_\tau \bar{\nu}_\tau$.   

The leading $\tan\beta$ corrections to the $Z\rightarrow b_R \bar{b}_R$ 
are shown in Figs.~\ref{neutralhiggs} and \ref{chargedhiggs}.
The amplitudes of these diagrams are:
\begin{eqnarray}
\lefteqn{-\frac{g^2}{4}
          \left( \frac{m_b\tan\beta}{m_W} \right)^2
          \left[ -i\frac{g}{\cos\theta_W} 
                 Z^\mu(Q)\,\bar{b}_R(p)\gamma_\mu b_R(q) 
          \right] \times} \qquad\qquad\qquad & & \cr
(2a+2b/h^0)
& : & \sin^2\!\alpha\,2\hat{C}_{24}(0,0,Q^2;0,m_{h^0},m_{A^0}) \cr
(2a+2b/H^0)
& : & \cos^2\!\alpha\,2\hat{C}_{24}(0,0,Q^2;0,m_{H^0},m_{A^0}) \cr
(2c/h^0)
& : & h_{b_L}\sin^2\!\alpha
      \left\{ (d-2)\hat{C}_{24}(0,0,Q^2;m_{h^0},0,0) \right.  \cr
&   & \qquad\qquad\qquad
      \left. - Q^2 \hat{C}_{23}(0,0,Q^2;m_{h^0},0,0) \right\} \cr
(2c/H^0)
& : & h_{b_L}\cos^2\!\alpha
      \left\{ (d-2)\hat{C}_{24}(0,0,Q^2;m_{H^0},0,0) \right.  \cr
&   & \qquad\qquad\qquad
      \left. - Q^2 \hat{C}_{23}(0,0,Q^2;m_{H^0},0,0) \right\} \cr
(2c/A^0)
& : & h_{b_L} \left\{ (d-2)\hat{C}_{24}(0,0,Q^2;m_{A^0},0,0) \right. \cr
&   & \qquad\quad
              \left.  -Q^2 \hat{C}_{23}(0,0,Q^2;m_{A^0},0,0) \right\} \cr
(2d+2e/h^0)
& : & 2h_{b_R}\sin^2\!\alpha\, B_1(0;0,m_{h^0}) \cr
(2d+2e/H^0)
& : & 2h_{b_R}\cos^2\!\alpha\, B_1(0;0,m_{H^0}) \cr
(2d+2e/A^0)
& : & 2h_{b_R}B_1(0;0,m_{A^0}) \cr
(3a)
& : & -4h_{H^+}\hat{C}_{24}(0,0,Q^2;m_t,m_{H^\pm},m_{H^\pm}) \cr
(3b)
& : & 2h_{t_L}\left\{ (d-2)\hat{C}_{24}(0,0,Q^2;m_{H^\pm},m_t,m_t)
              \right. \cr
&   & \qquad\quad\left.
                    -Q^2\hat{C}_{23}(0,0,Q^2;m_{H^\pm},m_t,m_t)
              \right\} \cr
(3c)
& : & -2h_{t_R} m_t^2 \hat{C}_0 (0,0,Q^2;m_{H^\pm},m_t,m_t) \cr
(3d+3e)
& : & 4h_{b_R} B_1(0;m_t,m_{H^\pm})
\end{eqnarray}
where
\begin{equation}
h_f = I_{3f} - Q_f \sin^2{\theta_W}
\end{equation}
and $Q^2=m_Z^2$.
The tree--level amplitude is the expression in the square brackets
times $h_{b_R}$.  As before, we have dropped terms subleading in $\tan\beta$.
Combining these corrections, with factors of $1/2$ for the wave--function
renormalizations, leads to a shift in the right--handed coupling of the
$b$ to the $Z$ given by
\begin{equation}
\delta h_{b_R} = \delta h_{b_R}^{N} + \delta h_{b_R}^{C}
\end{equation}
where
\begin{eqnarray}
\delta h_{b_R}^{N}
& = & -\frac{g^2}{4}\left( \frac{m_b\tan\beta}{m_W} \right)^2 \times \cr
&   & \Biggl[\,\sin^2\!\alpha
             \biggl\{ 2\hat{C}_{24}(0,m_{h^0},m_{A^0})
                   + \frac{1}{2}B_1(0,m_{h^0})
                   + \frac{1}{2}B_1(0,m_{A^0})
             \biggr\}
      \Biggr. \cr
&   &      + \cos^2\!\alpha
             \biggl\{ 2\hat{C}_{24}(0,m_{H^0},m_{A^0})
                   + \frac{1}{2}B_1(0,m_{H^0})
                   + \frac{1}{2}B_1(0,m_{A^0})
             \biggr\} \cr
&   &      + h_{b_L}\sin^2\!\alpha
             \biggl\{ (d-2)\hat{C}_{24}(m_{h^0},0,0)
                     -m_Z^2\hat{C}_{23}(m_{h^0},0,0) + B_1(0,m_{h^0})
             \biggr\} \cr
&   &      + h_{b_L}\cos^2\!\alpha
             \biggl\{ (d-2)\hat{C}_{24}(m_{H^0},0,0)
                     -m_Z^2\hat{C}_{23}(m_{H^0},0,0) + B_1(0,m_{H^0})
             \biggr\} \cr
&   & \Biggl.
           + h_{b_L}
             \biggl\{ (d-2)\hat{C}_{24}(m_{A^0},0,0)
                     -m_Z^2\hat{C}_{23}(m_{H^0},0,0) + B_1(0,m_{A^0})
             \biggr\}
      \Biggr], \cr
& & \cr
\delta h_{b_R}^{C}
& = & -\frac{g^2}{2}\left( \frac{m_b\tan\beta}{m_W} \right)^2 \times \cr
&   & \Biggl[ - h_{H^+}
             \biggl\{ 2\hat{C}_{24}(m_t,m_{H^\pm},m_{H^\pm})
                   +  B_1(m_t,m_{H^\pm})
             \biggr\}
      \Biggr. \cr
&   &      + h_{t_L}
             \biggl\{ (d-2)\hat{C}_{24}(m_{H^\pm},m_t,m_t)
                    - m_Z^2\hat{C}_{23}(m_{H^\pm},m_t,m_t)
                    + B_1(m_t,m_{H^\pm})
             \biggr\} \cr
&   & \Biggl.
           - h_{t_R} m_t^2\hat{C}_0(m_{H^\pm},m_t,m_t)
      \Biggr]
\end{eqnarray}
As in the $W$ decay case, these expressions can be well approximated by
their leading terms in an expansion in $m_Z^2$ as long as the
mass splittings among the Higgses are not small.  Using the formulae
from the previous section and from the Appendix, we find
\begin{eqnarray}
\delta h_{b_R}^{N}
& \approx & +\frac{1}{(4\pi)^2} \frac{g^2}{8}
             \left( \frac{m_b\tan\beta}{m_W} \right)^2
             \left[ \sin^2\!\alpha\;
                    G\!\left( \frac{m_{h^0}^2}{m_{A^0}^2} \right)
                  + \cos^2\!\alpha\;
                    G\!\left( \frac{m_{H^0}^2}{m_{A^0}^2} \right)
             \right] \cr
\delta h_{b_R}^{C}
& \approx & +\frac{1}{(4\pi)^2} \frac{g^2}{4}
             \left( \frac{m_b\tan\beta}{m_W} \right)^2
             F\!\left( \frac{m_t^2}{m_{H^\pm}^2} \right)
\end{eqnarray}
where the function $G(x)$ was defined in Eq.~(\ref{eq:Gdef}) and
\begin{equation}
F(x) = \frac{x}{1-x} \left( 1+ \frac{1}{1-x} \ln{x}\right).
\label{eq:Fdef}
\end{equation}
See appendix for details.

The diagrams which correct the decay $Z\rightarrow b_L\bar{b}_L$ is
the same as those shown in Figs.~\ref{neutralhiggs} and \ref{chargedhiggs}
with the replacements $b_R \leftrightarrow b_L$ and $t_L \leftrightarrow t_R$.
The amplitudes of the neutral Higgs diagrams are
\begin{eqnarray}
\lefteqn{+\frac{g^2}{4}
          \left( \frac{m_b\tan\beta}{m_W} \right)^2
          \left[ -i\frac{g}{\cos\theta_W} 
                 Z^\mu(Q)\,\bar{b}_L(p)\gamma_\mu b_L(q) 
          \right] \times} \qquad\qquad\qquad & & \cr
(2a+2b/h^0)
& : & \sin^2\!\alpha\,2\hat{C}_{24}(0,0,Q^2;0,m_{h^0},m_{A^0}) \cr
(2a+2b/H^0)
& : & \cos^2\!\alpha\,2\hat{C}_{24}(0,0,Q^2;0,m_{H^0},m_{A^0}) \cr
(2c/h^0)
& : & -h_{b_R}\sin^2\!\alpha\,\left\{(d-2)\hat{C}_{24}(0,0,Q^2;m_{h^0},0,0)- Q^2 \hat{C}_{23}(0,0,Q^2;m_{h^0},0,0)\right\} \cr
(2c/H^0)
& : & -h_{b_R}\cos^2\!\alpha\,\left\{(d-2)\hat{C}_{24}(0,0,Q^2;m_{H^0},0,0)-Q^2 \hat{C}_{23}(0,0,Q^2;m_{H^0},0,0)\right\}
\cr
(2c/A^0)
& : & -h_{b_R}\left\{(d-2)\hat{C}_{24}(0,0,Q^2;m_{A^0},0,0) -Q^2 \hat{C}_{23}(0,0,Q^2;m_{A^0},0,0)\right\}\cr
(2d+2e/h^0)
& : & -2h_{b_L}\sin^2\!\alpha\, B_1(0;0,m_{h^0}) \cr
(2d+2e/H^0)
& : & -2h_{b_L}\cos^2\!\alpha\, B_1(0;0,m_{H^0}) \cr
(2d+2e/A^0)
& : & -2h_{b_L}B_1(0;0,m_{A^0})
\end{eqnarray}
with $Q^2=m_Z^2$.
The charged Higgs diagrams lead to corrections proportional to
\[
\left( \frac{m_t\cot\beta}{m_W} \right)^2
\]
and are suppressed compared to the neutral Higgs diagrams by a factor of
$(m_t/m_b\tan^2\beta)^2 \sim (7.6/\tan\beta)^4$ so will be neglected.
The shift in the left--handed coupling of the $b$ to the $Z$ is then
\begin{equation}
\delta h_{b_L} = \delta h_{b_L}^{N} + \delta h_{b_L}^{C}
\end{equation}
with
\begin{eqnarray}
\delta h_{b_L}^{N}
& = & +\frac{g^2}{4}\left( \frac{m_b\tan\beta}{m_W} \right)^2 \times \cr
&   & \Biggl[\,\sin^2\!\alpha
             \biggl\{ 2\hat{C}_{24}(0,m_{h^0},m_{A^0})
                   + \frac{1}{2}B_1(0,m_{h^0})
                   + \frac{1}{2}B_1(0,m_{A^0})
             \biggr\}
      \Biggr. \cr
&   &      + \cos^2\!\alpha
             \biggl\{ 2\hat{C}_{24}(0,m_{H^0},m_{A^0})
                   + \frac{1}{2}B_1(0,m_{H^0})
                   + \frac{1}{2}B_1(0,m_{A^0})
             \biggr\} \cr
&   &      - h_{b_R}\sin^2\!\alpha
             \biggl\{ (d-2)\hat{C}_{24}(m_{h^0},0,0) - m_Z^2 \hat{C}_{23}(m_{h^0},0,0) + B_1(0,m_{h^0})
             \biggr\} \cr
&   &      - h_{b_R}\cos^2\!\alpha
             \biggl\{ (d-2)\hat{C}_{24}(m_{H^0},0,0) - m_Z^2 \hat{C}_{23}(m_{H^0},0,0) + B_1(0,m_{H^0})
             \biggr\} \cr
&   &      - h_{b_R}
             \biggl\{ (d-2)\hat{C}_{24}(m_{A^0},0,0) - m_Z^2 \hat{C}_{23}(m_{A^0},0,0) + B_1(0,m_{A^0})
             \biggr\} \cr
&   & \cr
\delta h_{b_L}^{C} & = & 0
\end{eqnarray}
Again, in the approximation $Q^2\rightarrow 0$, we find
\begin{eqnarray}
\delta h_{b_L}^{N}
& \approx & -\frac{1}{(4\pi)^2} \frac{g^2}{8}
             \left( \frac{m_b\tan\beta}{m_W} \right)^2
             \left[ \sin^2\!\alpha\;
                    G\!\left( \frac{m_{h^0}^2}{m_{A^0}^2} \right)
                  + \cos^2\!\alpha\;
                    G\!\left( \frac{m_{H^0}^2}{m_{A^0}^2} \right)
             \right] \cr
& = & -\delta h_{b_R}^{N}.
\end{eqnarray}
So in this approximation, the shift in the left--handed coupling of 
the $b$ quark due to neutral
Higgses is equal in magnitude but opposite in sign to the shift in
the right--handed coupling.

To estimate the corrections to $Z\rightarrow u \bar{u}, c \bar{c}$, we note that
the Higgs couplings to $u$ and $c$ quarks are suppressed either by $\tan\beta$ or
by small $d$ and $s$ quark masses.  Thus, we neglect these corrections.

The corrections to the $\tau$ couplings to the $Z$ can be obtained from
those of the $b$ couplings by the simple substitutions
\begin{eqnarray}
m_t,m_b         & \rightarrow & 0,m_\tau \cr
h_{t_L},h_{t_R} & \rightarrow & h_{\nu_L},0 \cr
h_{b_L},h_{b_R} & \rightarrow & h_{\tau_L}, h_{\tau_R}
\end{eqnarray}
which lead to
\begin{eqnarray}
\delta h_{\tau_R}^{N} \;=\; -\delta h_{\tau_L}^{N}
& \approx & +\frac{1}{(4\pi)^2} \frac{g^2}{8}
             \left( \frac{m_\tau\tan\beta}{m_W} \right)^2
             \left[ \sin^2\!\alpha\;
                    G\!\left( \frac{m_{h^0}^2}{m_{A^0}^2} \right)
                  + \cos^2\!\alpha\;
                    G\!\left( \frac{m_{H^0}^2}{m_{A^0}^2} \right)
             \right] \cr
\delta h_{\tau_R}^{C} \;=\; \phantom{-}\delta h_{\tau_L}^{C}
& = & 0.
\end{eqnarray}
Note that the charged Higgs contribution is zero since
$m_t$ is replaced by $m_\nu = 0$ and 
$F(m_\nu^2/m_{H^\pm}^2) = F(0) = 0$.

The decay $Z\rightarrow \nu_\tau\bar{\nu}_\tau$ is corrected by the
diagrams shown in Fig.~\ref{neutrinodiagrams}.
The amplitude of these diagrams are
\begin{eqnarray}
\lefteqn{-\frac{g^2}{2}
          \left( \frac{m_\tau\tan\beta}{m_W} \right)^2
          \left[ -i\frac{g}{\cos\theta_W}
                 Z^\mu(Q)\,\bar{\nu}_{\tau L}(p)\gamma_\mu
                                  \nu_{\tau L}(q)
          \right] \times} \qquad\qquad & & \cr
(4a) 
& : & h_{\tau_R} \left\{ (d-2)\hat{C}_{24}(0,0,Q^2;m_{H^\pm},0,0) 
                 \right. \cr
&   & \left.\qquad\quad -Q^2\hat{C}_{23}(0,0,Q^2;m_{H^\pm},0,0) 
      \right\} \cr
(4b)
& : & h_{H^+} 2\hat{C}_{24}(0,0,Q^2;0,m_{H^\pm},m_{H^\pm}) \cr
(4c+4d)
& : & 2h_{\nu_L} B_1(0;0,m_{H^\pm})
\end{eqnarray}
with $Q^2=m_Z^2$,
resulting in a shift of the neutrino coupling by
\begin{eqnarray}
\delta h_{\nu_L}^{C}
& = & -\frac{g^2}{2}\left( \frac{m_\tau\tan\beta}{m_W} \right)^2 \times \cr
&   & \Biggl[ h_{H^\pm}
             \biggl\{ 2\hat{C}_{24}(0,m_{H^\pm},m_{H^\pm})
                   +  B_1(0,m_{H^\pm})
             \biggr\}
      \Biggr. \cr
&   & \Biggl. + h_{\tau_R}
             \biggl\{ (d-2)\hat{C}_{24}(m_{H^\pm},0,0)
                    - m_Z^2\hat{C}_{23}(m_{H^\pm},0,0)
                    + B_1(0,m_{H^\pm})
             \biggr\}
      \Biggr]
\end{eqnarray}
As a consequence of $F(0)=0$ we find
\begin{equation}
\delta h_{\nu_L}^{C} = 0.
\end{equation}

To summarize, we have found that the non--zero shifts in the
fermion couplings in our approximation ($Q^2=0$) are:
\begin{eqnarray}
\delta h_{b_R}^{N}
\;=\; -\delta h_{b_L}^{N}
& = & +\frac{1}{(4\pi)^2} \frac{g^2}{8}
       \left( \frac{m_b\tan\beta}{m_W} \right)^2
       \left[ \sin^2\!\alpha\;G\!\left( \frac{m_{h^0}^2}{m_{A^0}^2} \right)
            + \cos^2\!\alpha\;G\!\left( \frac{m_{H^0}^2}{m_{A^0}^2} \right)
       \right] \cr
\delta h_{b_R}^{C}
& = & +\frac{1}{(4\pi)^2} \frac{g^2}{4}
       \left( \frac{m_b\tan\beta}{m_W} \right)^2
       F\!\left( \frac{m_t^2}{m_{H^\pm}^2} \right) \cr
\delta h_{\tau_R}^{N}
\;=\; -\delta h_{\tau_L}^{N}
& = & \left( \frac{m_\tau^2}{m_b^2} \right) \delta h_{b_R}^{N}.
\end{eqnarray}
Since $G(x)$ is negative semi--definite, the shifts in the
left--handed couplings of the $b$ and the $\tau$ due to the
neutral Higgs sector are both always {\it positive} while the shifts in 
the right--handed couplings are always {\it negative}, and they are all
proportional to the same linear combination of $G$--functions.
Also, since $-1 \le F(x) \le 0$, the
charged Higgs sector produces only a {\it negative} shift in 
$h_{b_R}$ of magnitude at most
\[
\frac{1}{(4\pi)^2}\frac{g^2}{4}
\left( \frac{m_b\tan\beta}{m_W} \right)^2.
\]
%

%%%%%%%%%%%%%%%%%%%%%%%%%%%%%%%%%%%%%%%%%%%%%%%%%%%%%%%%%%%%%%%%%%%%%%%%%%%%%%

\subsection{Fit to the data}
\label{sec:Fit}

We have identified the relevant vertex corrections to $Z$ decay
in the large $\tan\beta$ 2HDM. 
Using the LEP/SLD data to constrain their sizes will let us constrain 
the ratios
\[
\frac{m_{h^0}^2}{m_{A^0}^2},\quad
\frac{m_{H^0}^2}{m_{A^0}^2},\quad\mbox{and}\quad
\frac{m_t^2}{m_{H^\pm}^2}.
\]
All the neutral Higgs corrections are proportional to each other, so we
will use $\delta h_{\tau_L}^{N}$ as the fit parameter.  For the
charged Higgs correction we will use $\delta h_{b_R}^{C}$.

In addition to the proper vertex corrections, the 2HDM corrects $Z$ decay
through oblique corrections which can be expressed as corrections to
the $\rho$--parameter and the effective value of $\sin^2\theta_W$.
Since we will consider only ratios of partial widths and asymmetry
parameters in our fit, the $\rho$--parameter drops out from our analysis
and we need only consider the shift in $\sin^2\theta_W$ which we will
denote $\delta s^2$.\footnote{Similar techniques for isolating oblique
corrections into one or just a few phenomenological parameters to
extract constraints on proper vertex corrections were used in
Refs.~\cite{Lebedev:1999vc,Lebedev:1999ze,Loinaz:1999jg,Takeuchi:1994zh}.}
We will not utilize $\delta s^2$ to extract
information on the 2HDM because oblique corrections are generically
sensitive to other sorts of new physics as well.

The shifts to the $Zf\bar{f}$ couplings in the large $\tan\beta$
2HDM can then be expressed as:
\begin{eqnarray}
\delta h_{\nu_{eL}}     \;=\;
\delta h_{\nu_{\mu L}}  \;=\;    
\delta h_{\nu_{\tau L}} & = & 0        \cr
\delta h_{e_{L,R}}   \;=\; 
\delta h_{\mu_{L,R}} & = & \delta s^2 \cr
\delta h_{\tau_L}    & = & \delta s^2
                           + \delta h_{\tau_L}^{N} \cr
\delta h_{\tau_R}    & = & \delta s^2
                           - \delta h_{\tau_L}^{N} \cr
\delta h_{u_{L,R}}   \;=\; 
\delta h_{c_{L,R}}   & = & -\frac{2}{3}\delta s^2 \cr
\delta h_{d_{L,R}}   \;=\; 
\delta h_{s_{L,R}}   & = &  \frac{1}{3}\delta s^2 \cr
\delta h_{b_L} & = & \frac{1}{3}\delta s^2
                    + \left( \frac{m_b^2}{m_\tau^2} \right)
                      \delta h_{\tau_L}^{N} \cr
\delta h_{b_R} & = & \frac{1}{3}\delta s^2 
                    - \left( \frac{m_b^2}{m_\tau^2} \right)
                      \delta h_{\tau_L}^{N}
                    + \delta h_{b_R}^{C}
\label{eq:Zcouplingshifts}
\end{eqnarray}
The dependence of various observables on $\delta h_{b_R}^{N}$,
$\delta h_{b_R}^{C}$, and $\delta s^2$
can be calculated in a straightforward manner.  For example:
\begin{eqnarray*}
\frac{ \delta A_e }
     { A_e }
& = & \frac{ 4 h_{e_L} h_{e_R}
             ( h_{e_R}\delta h_{e_L} - h_{e_L}\delta h_{e_R} ) }
           { ( h_{e_L}^4 - h_{e_R}^4 }  \cr
& = & \frac{ 4 h_{e_L} h_{e_R} ( h_{e_R} - h_{e_L} ) }
           { h_{e_L}^4 - h_{e_R}^4 }
             \delta s^2 \cr
& = & -53.5\,\delta s^2 
\end{eqnarray*}
where the coefficient has been calculated assuming $\sin^2\theta_W = 0.2315$.
Similarly,
\begin{eqnarray}
\frac{ \delta A_\tau }{ A_\tau } 
& = & -53.5\,\delta s^2
      +3.96\,\delta h_{\tau_L}^{N} \cr
\frac{ \delta A_{\rm FB}(e) }{ A_{\rm FB}(e) } \;=\;
\frac{ \delta A_{\rm FB}(\mu) }{ A_{\rm FB}(\mu) }
& = & -107\,\delta s^2  \cr
\frac{ \delta A_{\rm FB}(\tau) }{ A_{\rm FB}(\tau) }
& = & -107\,\delta s^2
      +3.96\,\delta h_{\tau_L}^{N} \cr
\frac{ \delta R_e }{ R_e } \;=\;
\frac{ \delta R_\mu }{ R_\mu } 
& = & -0.84\,\delta s^2
      -2.89\,\delta h_{\tau_L}^{N}
      +0.184\,\delta h_{b_R}^{C}
      +0.307\,\delta\alpha_s   \cr
\frac{ \delta R_\tau }{ R_\tau }
& = & -0.84\,\delta s^2
      +5.07\,\delta h_{\tau_L}^{N}
      +0.184\,\delta h_{b_R}^{C}
      +0.307\,\delta\alpha_s \cr
\frac{ \delta R_b }{ R_b } 
& = & 0.182\,\delta s^2
     -10.3\,\delta h_{\tau_L}^{N}
     +0.652\,\delta h_{b_R}^{C} \cr
\frac{ \delta R_c }{ R_c }
& = & -0.351\,\delta s^2
      +2.89\,\delta h_{\tau_L}^{N}
      -0.184\,\delta h_{b_R}^{C} \cr
\frac{ \delta A_{\rm FB}(b) }{ A_{\rm FB}(b) }
& = & -54.1\,\delta s^2
      +3.43\,\delta h_{\tau_L}^{N}
      -1.73\,\delta h_{b_R}^{C} \cr
\frac{ \delta A_{\rm FB}(c) }{ A_{\rm FB}(c) }
& = & -58.7\,\delta s^2 \cr
\frac{ \delta A_b }{ A_b }
& = & -0.681\,\delta s^2
      +3.43\,\delta h_{\tau_L}^{N}
      -1.73\,\delta h_{b_R}^{C} \cr
\frac{ \delta A_c }{ A_c }
& = & -5.19\,\delta s^2
\label{eq:FITCOEFFS} 
\end{eqnarray}
for $\sin^2\theta_W = 0.2315$, $\overline{m}_\tau(m_Z) = 1.777$~GeV, and 
$\overline{m}_b(m_Z) = 2.77$~GeV.
\footnote{Letting $\overline{m}_b(m_Z)$ change by 10\% in either direction
will only change the final central value of $\delta h_{\tau_L}^{N}$
by about a fifth of a $\sigma$.}
We have introduced the parameter $\delta\alpha_s$
to account for the deviation of $\alpha_s(m_Z)$ from its
nominal value which we chose to be 0.120\footnote{%
The value of $\alpha_s(m_Z)$ from LEP is determined by fitting the SM to 
$R_\ell$, ($\ell=e,\mu,\tau$).  Since we are considering extra corrections
to the $R_\ell$'s, we must let $\alpha_s(m_Z)$ float in our fit.}:
\[
\alpha_s(m_Z) = 0.120 + \delta\alpha_s.
\]
We fit the expressions in Eq.~(\ref{eq:FITCOEFFS}) to the differences 
between the LEP/SLD measurements and SM predictions shown in 
Table~\ref{LEP-SLD-DATA}.
The corresponding correlation matrices of the data are given in 
Tables~\ref{lineshape-correlations} and \ref{heavy-correlations}.  
The SM predictions listed are for a SM Higgs mass of 300~GeV.
Changing the SM Higgs mass has a negligible effect on all fit parameters 
except $\delta s^2$ which,
as discussed above, we do not utilize except as a fit parameter.

The result of the fit was
\begin{eqnarray}
\delta h_{\tau_L}^{N} & = & -0.00021 \pm 0.00029 \cr
\delta h_{b_R}^{C}    & = & \phantom{-}0.0049 \pm 0.0060 \cr
\delta s^2     & = & -0.00069 \pm 0.00019 \cr
\delta\alpha_s & = & -0.0007 \pm 0.0051
\label{eq:FITRESULT}
\end{eqnarray}
with the correlation matrix for the fit parameters shown in 
Table~\ref{fit-correlation}.
The quality of the fit was $\chi^2 = 18.4/(18-4)$. 
The largest contributions to the $\chi^2$ come from
$A_{\rm FB}(b)$ (3.5) and $A_{\rm LR}$ (2.5) which means that
the 2HDM corrections do not improve the agreement between the
theoretical and experimental values of these observables.

In Figs.~\ref{Contour1}, \ref{Contour2}, and \ref{Contour3} 
we show how different observables constrain the
parameters $\delta h_{\tau_L}^{N}$, $\delta h_{b_R}^{C}$, and
$\delta s^2$.
Since $\delta h_{\tau_L}^{N}$ is the only parameter which breaks lepton
universality, it is most strongly constrained by the ratios
$R_\ell$ ($\ell = e,\mu,\tau$).
This is evident from Figs.~\ref{Contour1} and \ref{Contour2}.
This places a tight constraint on the size of the neutral Higgs 
correction to the $b$ quark observables.\footnote{This was pointed 
out by Hisano et al. in Ref.~\cite{Hisano:1997fx}.}
The charged Higgs contribution, $\delta h_{b_R}^{C}$, must then fit 
all the heavy flavor observables, but due to the small experimental
error on $R_b$, it is also constrained to be small.
In Fig.~\ref{Contour3}, one sees that the
overlap of the $A_{\rm LR}$ and $A_{\rm FB}(b)$ bands
prefers a value of $\delta h_{b_R}$ of about 0.04, far from the
SM point at the origin.\footnote{This is the `$A_b$ anomaly' mentioned 
in the introduction.  
See, for instance, Refs.~\cite{Chanowitz:1999jj} and \cite{Field:1999ng}.}
However, the $R_b$ band does not allow this deviation, leading to
the large $\chi^2$'s for $A_{\rm LR}$ and $A_{\rm FB}(b)$ mentioned
above.
Note also that the large $\tan\beta$ 2HDM predicts 
$\delta h_{b_R}^{C} \leq 0$ so it cannot account for the 
`$A_b$ anomaly' even if the constraint from $R_b$ were absent.
In fact, since the best fit value of $\delta h_{b_R}^{C}$ is still
small but positive, the 2HDM is slightly disfavored by the data.

The same can be said of $\delta h_{\tau_L}^{N}$:
Since the experimental value of $R_\tau$ is smaller than those
for $R_\mu$ and $R_e$, it is easy to see from Eq.~(\ref{eq:FITCOEFFS}) that
the data prefer a negative value of $\delta h_{\tau_L}^{N}$.
However, the large $\tan\beta$ 2HDM predicts $\delta h_{\tau_L}^{N} \geq 0$.

%%%%%%%%%%%%%%%%%%%%%%%%%%%%%%%%%%%%%%%%%%%%%%%%%%%%%%%%%%%%%%%%%%%%%%%%%%%%%%

\subsection{Constraints on model parameters}

Since $\delta h_{\tau_L}^{N} \geq 0$ and $\delta h_{b_R}^{C} \leq 0$ 
for the large $\tan\beta$ 2HDM in our approximation,
the best fit values of $\delta h_{\tau_L}^{N}$ and $\delta h_{b_R}^{C}$ 
consistent with the model are 
$\delta h_{\tau_L}^{N} = \delta h_{b_R}^{C} = 0$, 
corresponding to the Standard Model case.
Thus, the large $\tan\beta$ 2HDM does not mitigate the $A_b$ problem, 
nor does it even improve agreement between the theoretical predictions for 
$R_b$ and $R_\tau$ and the experimental data.\footnote{Similar behavior in the 
context of the MSSM with R--parity violation was observed in 
Ref.~\cite{Lebedev:1999ze}.  There, the preferred values of the fit parameters
were again the opposite sign of what the model predicted, and moreover,
more than one to two $\sigma$ away from zero.
This was a manifestation of the $A_b$ anomaly.
In the model considered here, the $A_b$ anomaly is not as manifest in
the fit results, since there are fewer new physics parameters.  
}

In order to extract the limits on the Higgs mass ratios 
from Eq.~(\ref{eq:FITRESULT}), we have performed both classical and 
Bayesian statistical analyses, in the latter assuming a uniform
prior probability for the parameter regions
$\delta h_{\tau_L}^{N}\geq 0$ and 
\begin{equation}
0
\geq \delta h_{b_R}^{C} 
\geq -\frac{1}{(4\pi)^2}\frac{g^2}{4}
      \left( \frac{m_b\tan\beta}{m_W} \right)^2
\end{equation}
(Recall that
\[
\delta h_{b_R}^{C} 
= \frac{1}{(4\pi)^2} \frac{g^2}{4}
  \left( \frac{m_b\tan\beta}{m_W} \right)^2
  F\!\left( \frac{m_t^2}{m_{H^\pm}^2} \right)
\]
and $-1 \le F(x) \le 0$.)

The corresponding \{68\%\} and [95\%] confidence limits on the 
fit parameters are:
\begin{eqnarray}
\mbox{classical} & : &
\delta h_{b_R}^{C} \;\geq\; \{\, -0.0011 \,\} \; [\, -0.0071 \,] \cr
& &
\delta h_{\tau_L}^{N} \;\leq\; \{\, 0.00008 \,\} \; [\, 0.00037 \,] \cr
& & \cr
\mbox{Bayesian} & : &
\delta h_{b_R}^{C} \;\geq\; \{\, -0.0021 \,\} \; [\, -0.0050 \,] \cr
& &
\delta h_{\tau_L}^{N} \;\leq\; \{\, 0.00011 \,\} \; [\, 0.00025 \,] 
\end{eqnarray}
Using $2m_W/g=246\;{\rm GeV}$ and ${\overline m}_b(m_Z)=2.77\;{\rm GeV}$,
the bounds on $\delta h_{b_R}^{C}$ translate into the bounds on $F(x)$, 
where $x=m_t^2/ m_{H^\pm}^2$:
\begin{eqnarray}
\mbox{classical} & : &
F(x) \;\geq\;
\left\{ -\left(\frac{37}{\tan\beta}\right)^2 \right\} \;
\left[  -\left(\frac{94}{\tan\beta}\right)^2 \right] \cr
\mbox{Bayesian}  & : & 
F(x) \;\geq\;
\left\{ -\left(\frac{52}{\tan\beta}\right)^2 \right\} \;
\left[  -\left(\frac{79}{\tan\beta}\right)^2 \right].
\end{eqnarray}
For $\tan\beta < 94$, the entire range of $F(x)$ is contained
in the classical 95\% confidence region (since $-1 \leq F(x) \leq 0$).
It is therefore difficult to significantly bound the charged Higgs mass 
using this method unless $\tan\beta$ is quite large.  
Choosing $\tan\beta=100$ for definiteness, 
we translate the bounds on $F(x)$ into bounds on $m_{H^\pm}$ 
\begin{eqnarray}
\mbox{classical} & : &
m_{H^\pm} \;\geq\; \{\, 670  \,{\rm GeV}\,\} \; [\, 40  \,{\rm GeV}\,] \cr
\mbox{Bayesian} & : &
m_{H^\pm} \;\geq\; \{\, 370  \,{\rm GeV}\,\} \; [\, 120  \,{\rm GeV}\,] 
\end{eqnarray}
Bounds on $m_{H^\pm}$ are plotted as a function of $\tan\beta$ in 
Fig.~\ref{chargedhiggsmass}.

Bounds on the neutral Higgs sector masses from constraints on 
$\delta h_{\tau_L}^{N}$ are more model--dependent
than charged--Higgs bounds, since $\delta h_{\tau_L}^{N}$ involves 
the masses of all of the neutral Higgses and the mixing angle $\alpha$.  
Generally, requiring the magnitude of $\delta h_{\tau_L}^N$ to be small 
constrains the scalar--pseudoscalar mass splittings to be small.  
To give a concrete example (as in Refs.~\cite{Hisano:1997fx,Haber:1999zh}), 
let us consider the limit $\alpha=\beta$ and $m_{A^0}=m_{H^0}$, 
\begin{equation}
\delta h_{\tau_L}^{N}
= - \frac{1}{(4\pi)^2} \frac{g^2}{8} 
\left( \frac{ m_\tau \tan\beta }{ m_W }
\right)^2 
G\!\left( \frac{ m_{h^0}^2 }{ m_{A^0}^2 } \right).
\end{equation}
In this approximation the 
\{68\%\} and [95\%] lower limits on 
$G\!\left(\frac{m_{h^0}^2}{m_{A^0}^2}\right)$ 
(which is negative semi--definite) are 
\begin{eqnarray}
\mbox{classical} & : &
G\!\left( \frac{ m_{h^0}^2 }{ m_{A^0}^2 } \right) \;\geq\;  
\left\{ -\left( \frac{22}{\tan\beta} \right)^2 \right\} \;
\left[  -\left( \frac{47}{\tan\beta} \right)^2 \right] \cr
\mbox{Bayesian} & : &
G\!\left( \frac{ m_{h^0}^2 }{ m_{A^0}^2 } \right) \;\geq\; 
\left\{ -\left( \frac{25}{\tan\beta} \right)^2 \right\} \;
\left[  -\left( \frac{39}{\tan\beta} \right)^2 \right].
\end{eqnarray}
With the choice $\tan\beta=100$ we find that our bounds on $G$ translate 
into bounds on the $h^0$--$A^0$ mass splitting 
(choosing the branch of solutions with $m_{h^0}/m_{A^0}<1$):
\begin{eqnarray}
\mbox{classical} & : &
1 \;\geq\; \frac{ m_{h^0} }{ m_{A^0} } \;\geq\; 
\left\{\,  0.68    \,\right\} \; 
\left[\,   0.43    \,\right] \cr
\mbox{Bayesian} & : &
1 \;\geq\; \frac{ m_{h^0} }{ m_{A^0} } \;\geq\; 
\left\{\,  0.64    \,\right\} \;
\left[\,   0.51    \,\right] 
\end{eqnarray}
More generally, bounds on $m_{h^0}/m_{A^0}$ are plotted as a 
function of $\tan\beta$ in Fig.~\ref{neutralhiggsmass}.
In the limit  $\sin\alpha \sim 0$,
the result is the same, except that $m_{H^0}/m_{A^0}$ replaces 
$m_{h^0}/m_{A^0}$ in the above expressions.

%%%%%%%%%%%%%%%%%%%%%%%%%%%%%%%%%%%%%%%%%%%%%%%%%%%%%%%%%%%%%%%%%%%%%%%%%%%%%%

\section{Summary and Conclusions}

We have analyzed the implications of the large $\tan\beta$ 2HDM for 
$Z$ decays and for lepton universality violation in $W$ decays.  
For $Z$ decays we find that the generic predictions of the 
model do not improve agreement between theory and experiment.  
Further, the LEP/SLD experimental uncertainty in the measurement of 
lepton universality is sufficiently small to place
significant constraints on mass splittings in the neutral Higgs sector 
for large $\tan\beta$.
Constraints from the $b$ decay parameters are sufficient to place 
bounds on the charged Higgs mass that are 
increasingly strong for increasing $\tan\beta$.
For instance, for $\tan\beta=100$ we obtain the
$1\sigma$ classical (Bayesian) bounds of
\begin{equation}
m_{H^\pm} \geq  670 \,{\rm GeV} \; (  370 \,{\rm GeV})
\quad\mbox{and}\quad
1 \geq \frac{ m_{h^0} }{ m_{A^0} } \geq 0.68   \; ( 0.64   ).
\end{equation}

For $W$ decays, the experimental central value from D{0\kern-6pt/} slightly 
disfavors the generic prediction of the model, but experimental uncertainties
are too large to usefully constrain the charged--neutral Higgs mass splittings.

%%%%%%%%%%%%%%%%%%%%%%%%%%%%%%%%%%%%%%%%%%%%%%%%%%%%%%%%%%%%%%%%%%%%%%%%%%%%
%\newpage

\acknowledgements

This work was supported in part (O.L. and W.L.) by the 
U.~S. Department of Energy, grant DE-FG05-92-ER40709, Task A.

%%%%%%%%%%%%%%%%%%%%%%%%%%%%%%%%%%%%%%%%%%%%%%%%%%%%%%%%%%%%%%%%%%%%%%%%%%%
%%%%%%%%%%%%%%%%%%%%%%%%%%%%%%%%%%%%%%%%%%%%%%%%%%%%%%%%%%%%%%%%%%%%%%%%%%%%

\onecolumn
\widetext

\appendix

\newcommand{\pole}{\Delta_\epsilon}

\section*{Feynman Integrals}
The integrals we use here are defined explicitly in \cite{Lebedev:1999vc}.
In the approximation $p^2=0$, the one-loop diagrams which appear in this
work are proportional to the following expressions:
\setlength{\unitlength}{1cm}
\begin{eqnarray}
\raisebox{-1cm}{\begin{picture}(2.5,2)(0,0)
\epsfbox[0 0 90 60]{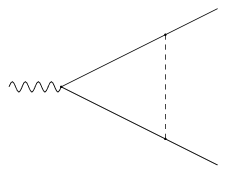}
\end{picture}}
& \propto & 
\left[ (d-2)\,\hat{C}_{24}\left( 0,0,p^2;m_{s},m_{f},m_{f}
                          \right)
     - m_Z^2\,\hat{C}_{23}\left( 0,0,p^2;m_{s},m_{f},m_{f}
                          \right)
\right] 
\cr
& \approx & 
-\frac{1}{(4\pi)^2} 
\left[ \frac{1}{2} \left( \pole - \ln{\frac{m_f^2}{\mu^2}} \right) + f(x)
\right] 
\\
& & \cr
\raisebox{-1cm}{\begin{picture}(2.5,2)(0,0)
\epsfbox[0 0 90 60]{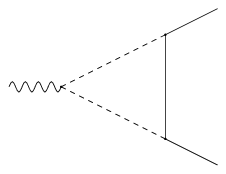}
\end{picture}}
& \propto & 
2\,\hat{C}_{24}\left( 0,0,p^2;m_{f},m_{s},m_{s} \right)
\;\approx\; 
- \frac{1}{(4\pi)^2}
\left[ \frac{1}{2} \left( \pole - \ln{\frac{m_f^2}{\mu^2}} \right) - g(x)
\right] 
\\
& & {\rm or} \nonumber \\ 
& \propto & 
2\,\hat{C}_{24}\left( 0,0,p^2;0,m_{s_1},m_{s_2} \right)
\;\approx\; 
- \frac{1}{2(4\pi)^2}
\left[  \pole  + \frac{3}{2}
      - \frac{ m_{s_1}^2 \ln{m_{s_1}^2} - m_{s_2}^2 \ln{m_{s_2}^2} }
             { m_{s_1}^2 - m_{s_2}^2 }
\right]
\\
& & \cr
\raisebox{-1cm}{\begin{picture}(2.5,2)(0,0)
\epsfbox[0 0 90 60]{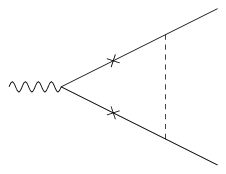}
\end{picture}}
& \propto &  
m_f^2\,\hat{C}_{0}\left( 0,0,p^2;m_{s},m_{f},m_{f} \right) 
\;\approx\; 
-\frac{1}{(4\pi)^2} \left[\, f(x)+ g(x) \,\right]
\\
& & \cr
\raisebox{-1cm}{\begin{picture}(2.5,2)(0,0)
\epsfbox[0 0 90 60]{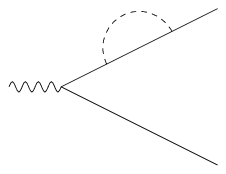}
\end{picture}}
& \propto &  
\hat{B}_{1} \left( 0;m_{f},m_{s} \right)
\;\approx\;
\frac{1}{(4\pi)^2}
\left[ \frac{1}{2} \left( \pole - \ln{\frac{m_f^2}{\mu^2}} \right) - g(x)
\right]
\end{eqnarray}
where 
\begin{eqnarray}
f(x) & = & -\frac{1}{4 (1-x)^2}\left(\,x^2 - 1 - 2 \ln{x} \,\right) \cr
g(x) & = & -\frac{1}{2} \ln{x} + \frac{1}{4 (1-x)^2}
            \left[ -(1-x)(1-3 x) + 2 x^2 \ln{x} \,\right]
\end{eqnarray}
for $x = m_f^2 / m_s^2.$  Note that the function $F(x)$ appearing in 
Eq.~(\ref{eq:Fdef}) is defined as
\[
F(x)=f(x)+g(x).
\]
For $x \longrightarrow 1$ (degenerate scalar and fermion masses),
\begin{eqnarray}
f(x) & \approx & -\frac{1}{2} + \frac{x-1}{6} + \cdots \cr 
g(x) & \approx & - \frac{x-1}{3} + \cdots
\end{eqnarray}
For $x \longrightarrow 0$ (the decoupling limit of heavy scalar masses),
\begin{eqnarray}
f(x) & \approx & \phantom{-}\frac{1}{2} \ln{x} + \frac{1}{4} + \cdots \cr 
g(x) & \approx & - \frac{1}{2} \ln{x}  -  \frac{1}{4} + \cdots 
     \;\sim\; -f(x).
\end{eqnarray}
The function $G(x)$ defined in Eq.~(\ref{eq:Gdef}) is symmetric under 
$x \leftrightarrow 1/x$.  For  $x \rightarrow 0,\infty$,
\begin{equation}
G(x) \sim -\biggl|\frac{\ln{x}}{2}\biggr|+1 + \cdots.
\end{equation}
For  $x \longrightarrow 1$ (the decoupling limit of degenerate scalar masses),
\begin{equation}
G(x) \sim -\frac{1}{12} (x-1)^2 +\cdots.
\end{equation}

%%%%%%%%%%%%%%%%%%%%%%%%%%%%%%%%%%%%%%%%%%%%%%%%%%%%%%%%%%%%%%%%%%%%%%%%%%
\widetext

\begin{table}[p]
\begin{center}
\begin{tabular}{|c|c|c|c|}
Observable & Reference & Measured Value & ZFITTER Prediction \\
\hline\hline
\multicolumn{2}{|l|}{\underline{$Z$ lineshape variables}} & & \\
$m_Z$                & \cite{MNICH:99} & $91.1872 \pm 0.0021$ GeV & input \\
$\Gamma_Z$           & \cite{MNICH:99} & $2.4944 \pm 0.0024$ GeV  & unused \\
$\sigma_{\rm had}^0$ & \cite{MNICH:99} & $41.544 \pm 0.037$ nb    & unused \\
$R_e$                & \cite{MNICH:99} & $20.803 \pm 0.049$       & $20.739$ \\
$R_\mu$              & \cite{MNICH:99} & $20.786 \pm 0.033$       & $20.739$ \\
$R_\tau$             & \cite{MNICH:99} & $20.764 \pm 0.045$       & $20.786$ \\
$A_{\rm FB}(e   )$   & \cite{MNICH:99} & $0.0145 \pm 0.0024$      & $0.0152$ \\
$A_{\rm FB}(\mu )$   & \cite{MNICH:99} & $0.0167 \pm 0.0013$      & $0.0152$ \\
$A_{\rm FB}(\tau)$   & \cite{MNICH:99} & $0.0188 \pm 0.0017$      & $0.0152$ \\
\hline
\multicolumn{2}{|l|}{\underline{$\tau$ polarization at LEP}} & & \\
$A_e$        & \cite{MNICH:99} & $0.1483 \pm 0.0051$   & $0.1423$ \\ 
$A_\tau$     & \cite{MNICH:99} & $0.1424 \pm 0.0044$   & $0.1424$ \\
\hline
\multicolumn{2}{|l|}{\underline{SLD left--right asymmetries}} & & \\
$A_{LR}$     & \cite{SLD:99} & $0.15108 \pm 0.00218$ & $0.1423$ \\
$A_e$        & \cite{SLD:99} & $0.1558  \pm 0.0064$  & $0.1423$ \\
$A_{\mu}$    & \cite{SLD:99} & $0.137   \pm 0.016$   & $0.1424$ \\
$A_{\tau}$   & \cite{SLD:99} & $0.142   \pm 0.016$   & $0.1424$ \\
\hline
%\multicolumn{2}{|l|}{\underline{light quark flavor}} & & \\
%$R_s^{\prime *}$  [OPAL]   & \cite{OPAL:97} & $0.392  \pm 0.062$  & $0.360$\\
%$A_{\rm FB}^*(s)$ [OPAL]   & \cite{OPAL:97} & $0.075  \pm 0.029$  & $0.100$\\
%$A_{\rm FB}^*(u)$ [OPAL]   & \cite{OPAL:97} & $0.086  \pm 0.037$  & $0.071$\\
%$A_{\rm FB}^{**}(s)$ [DELPHI]
%                           & \cite{DELPHI:99} & $0.1008\pm 0.0120$ & $0.1006$\\
%$A_s^{*}$ [SLD]            & \cite{As-SLD:99} & $0.85  \pm 0.092$  & $0.935$ \\
%\hline
\multicolumn{2}{|l|}{\underline{heavy quark flavor}} & & \\
$R_b$           & \cite{MNICH:99} & $0.21642 \pm 0.00073$ & $0.21583$ \\
$R_c$           & \cite{MNICH:99} & $0.1674  \pm 0.0038$  & $0.1722$ \\
$A_{\rm FB}(b)$ & \cite{MNICH:99} & $0.0988  \pm 0.0020$  & $0.0997$ \\
$A_{\rm FB}(c)$ & \cite{MNICH:99} & $0.0692  \pm 0.0037$  & $0.0711$ \\
$A_b$           & \cite{MNICH:99} & $0.911   \pm 0.025$   & $0.934$ \\
$A_c$           & \cite{MNICH:99} & $0.630   \pm 0.026$   & $0.666$ \\
\end{tabular}
\caption{LEP/SLD observables and their Standard Model predictions.
The Standard Model predictions were calculated using ZFITTER v.6.21 
\protect\cite{ZFITTER:99} with $m_t = 174.3$~GeV \protect\cite{TOPMASS:99},
$m_H = 300$~GeV, and $\alpha_s(m_Z) = 0.120$ as input.}
\label{LEP-SLD-DATA}
\end{center}
\end{table}

\medskip

\widetext

\begin{table}[ht]
\begin{center}
\begin{tabular}{|c|ccccccccc|}
& $m_Z$     & $\Gamma_Z$     & $\sigma_{\rm had}^0$
& $R_e$     & $R_\mu$     & $R_\tau$ 
& $A_{\rm FB}(e)$ & $A_{\rm FB}(\mu)$ & $A_{\rm FB}(\tau)$ \\
\hline
$m_Z$ 
& $1.000$            & $-0.008$           & $-0.050$ 
& $\phantom{-}0.073$ & $\phantom{-}0.001$ & $\phantom{-}0.002$
& $-0.015$           & $\phantom{-}0.046$ & $\phantom{-}0.034$ \\
$\Gamma_Z$
&                    & $\phantom{-}1.000$ & $-0.284$ 
& $-0.006$           & $\phantom{-}0.008$ & $\phantom{-}0.000$ 
& $-0.002$           & $\phantom{-}0.002$ & $-0.003$ \\
$\sigma_{\rm had}^0$
&                    &                    & $\phantom{-}1.000$
& $\phantom{-}0.109$ & $\phantom{-}0.137$ & $\phantom{-}0.100$ 
& $\phantom{-}0.008$ & $\phantom{-}0.001$ & $\phantom{-}0.007$ \\
$R_e$
&                    &                    &
& $\phantom{-}1.000$ & $\phantom{-}0.070$ & $\phantom{-}0.044$ 
& $-0.356$           & $\phantom{-}0.023$ & $\phantom{-}0.016$ \\
$R_\mu$
&                    &                    &
&                    & $\phantom{-}1.000$ & $\phantom{-}0.072$ 
& $\phantom{-}0.005$ & $\phantom{-}0.006$ & $\phantom{-}0.004$ \\
$R_\tau$
&                    &                    &
&                    &                    & $\phantom{-}1.000$
& $\phantom{-}0.003$ & $-0.003$           & $\phantom{-}0.010$ \\
$A_{\rm FB}(e)$
&                    &                    & 
&                    &                    &
& $\phantom{-}1.000$ & $-0.026$           & $-0.020$ \\
$A_{\rm FB}(\mu)$ 
&                    &                    & 
&                    &                    & 
&                    & $\phantom{-}1.000$ & $\phantom{-}0.045$ \\
$A_{\rm FB}(\tau)$
&                    &                    & 
&                    &                    & 
&                    &                    & $\phantom{-}1.000$ \\
\end{tabular}
\caption{The correlation of the $Z$ lineshape variables at LEP.}
\label{lineshape-correlations}
\end{center}
\end{table}

\begin{table}[ht]
\begin{center}
\begin{tabular}{|c|cccccc|}
& $R_b$              & $R_c$     
& $A_{\rm FB}(b)$    & $A_{\rm FB}(c)$     
& $A_b$              & $A_c$ \\
\hline
$R_b$ 
& $1.00$            & $-0.14$           & $-0.03$ 
& $\phantom{-}0.01$ & $-0.03$           & $\phantom{-}0.02$ \\
$R_c$
&                   & $\phantom{-}1.00$ & $\phantom{-}0.05$ 
& $-0.05$           & $\phantom{-}0.02$ & $-0.02$ \\
$A_{\rm FB}(b)$
&                   &                   & $\phantom{-}1.00$
& $\phantom{-}0.09$ & $\phantom{-}0.02$ & $\phantom{-}0.00$ \\
$A_{\rm FB}(c)$
&                   &                   &
& $\phantom{-}1.00$ & $-0.01$           & $\phantom{-}0.03$ \\
$A_b$
&                   &                   &
&                   & $\phantom{-}1.00$ & $\phantom{-}0.15$ \\ 
$A_c$
&                   &                   &
&                   &                   & $\phantom{-}1.00$ \\
\end{tabular}
\caption{The correlation of the heavy flavor variables from LEP/SLD.}
\label{heavy-correlations}
\end{center}
\end{table}

\begin{table}[ht]
\begin{center}
\begin{tabular}{|c|rrrr|}
& $\delta h_{\tau_L}^{N}$ 
& $\delta h_{b_R}^{C}$ 
& $\delta s^2$ 
& $\delta\alpha_s$  \\
\hline
$\delta h_{\tau_L}^{N}$ 
& $1.00$ & $0.62$ & $-0.12$ & $-0.30$ \\
$\delta h_{b_R}^{C}$
&        & $1.00$ & $-0.22$ & $-0.63$  \\
$\delta s^2$
&        &        &  $1.00$ &  $0.25$ \\
$\delta\alpha_s$
&        &        &         &  $1.00$ \\
\end{tabular}
\caption{The correlation matrix of the fit parameters.}
\label{fit-correlation}
\end{center}
\end{table}

%%%%%%%%%%%%%%%%%%%%%%%%%%%%%%%%%%%%%%%%%%%%%%%%%%%%%%%%%%%%%%%%%%%%%%%%%%%%%
\newpage
\widetext

\newpage

\begin{figure}[ht]
\begin{center}
\unitlength=1cm
\begin{picture}(17,6)(0,0)
\unitlength=1mm
%\zahyou{17}{6}
\put(13,31){\vector(1,0){8}}
\put(15,25){$Q$}
\put(40,47){\vector(2,1){6}}
\put(40,50){$p$}
\put(40,22){\vector(2,-1){6}}
\put(40,18){$q$}
\put(30,5){(a)}
\put(85,5){(b)}
\put(140,5){(c)}
\put(6,33){$W^-$}
\put(61,33){$W^-$}
\put(116,33){$W^-$}
\put(41,33){$\tau_R$}
\put(50,48){$\tau_L$}
\put(106,48){$\tau_L$}
\put(160,48){$\tau_L$}
\put(50,19){$\nu_\tau$}
\put(106,19){$\nu_\tau$}
\put(160,19){$\nu_\tau$}
\put(17,42){$h_0,H_0,A_0$}
\put(26,25){$H^+$}
\put(77,39){$\tau_L$}
\put(87,44){$\tau_R$}
\put(90,31){$h_0,H_0,A_0$}
\put(133,29){$\nu_\tau$}
\put(142,24){$\tau_R$}
\put(150,34){$H^+$}
\epsfbox[0 600 480 780]{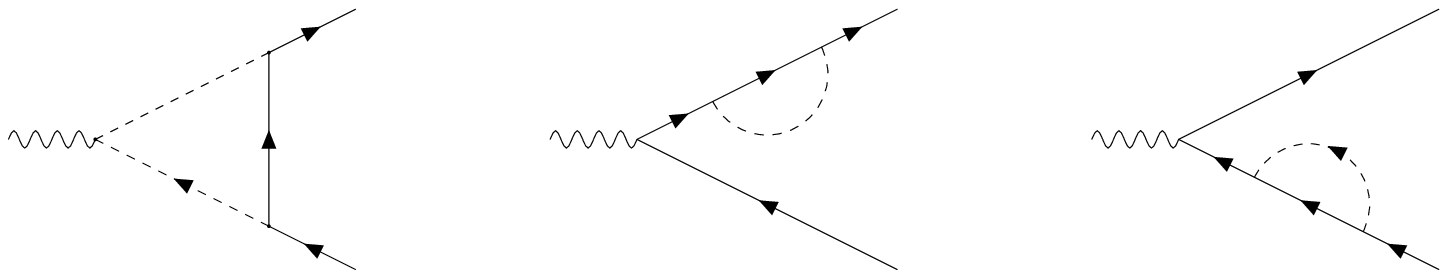}
\end{picture}
\caption{One--loop corrections to $W^- \rightarrow \tau_L 
{\overline\nu}_\tau$}
\label{Wdiagrams}
\end{center}
\end{figure}

%%%%%%%%%%%%%%%%%%%%%%%%%%%%%%%%%%%%%%%%%%%%%%%%%%%%%%%%%%%%%%%%%%%%%%%%%%%%%

\begin{figure}[ht]
\begin{center}
\unitlength=1cm
\begin{picture}(17,11)(0,0)
\unitlength=1mm
%\zahyou{17}{11}
\put(13,77){\vector(1,0){8}}
\put(15,71){$Q$}
\put(40,92){\vector(2,1){6}}
\put(40,95){$p$}
\put(40,68){\vector(2,-1){6}}
\put(40,64){$q$}
\put(30,55){(a)}
\put(85,55){(b)}
\put(140,55){(c)}
\put(60,5){(d)}
\put(115,5){(e)}
\put(8,79){$Z$}
\put(42,79){$b_L$}
\put(63,79){$Z$}
\put(97,79){$b_L$}
\put(118,79){$Z$}
\put(151,79){$h_0,H_0,A_0$}
\put(35,30){$Z$}
\put(90,30){$Z$}
\put(50,93){$b_R$}
\put(50,65){$b_R$}
\put(106,93){$b_R$}
\put(106,65){$b_R$}
\put(160,93){$b_R$}
\put(160,65){$b_R$}
\put(78,45){$b_R$}
\put(78,16){$b_R$}
\put(133,45){$b_R$}
\put(133,16){$b_R$}
\put(21,87){$h_0,H_0$}
\put(26,72){$A_0$}
\put(82,87){$A_0$}
\put(77,71){$h_0,H_0$}
\put(138,88){$b_L$}
\put(138,71){$b_L$}
\put(62,28){$h_0,H_0,A_0$}
\put(51,37){$b_R$}
\put(60,41){$b_L$}
\put(106,25){$b_R$}
\put(115,20){$b_L$}
\put(119,32){$h_0,H_0,A_0$}
\epsfbox[0 470 480 780]{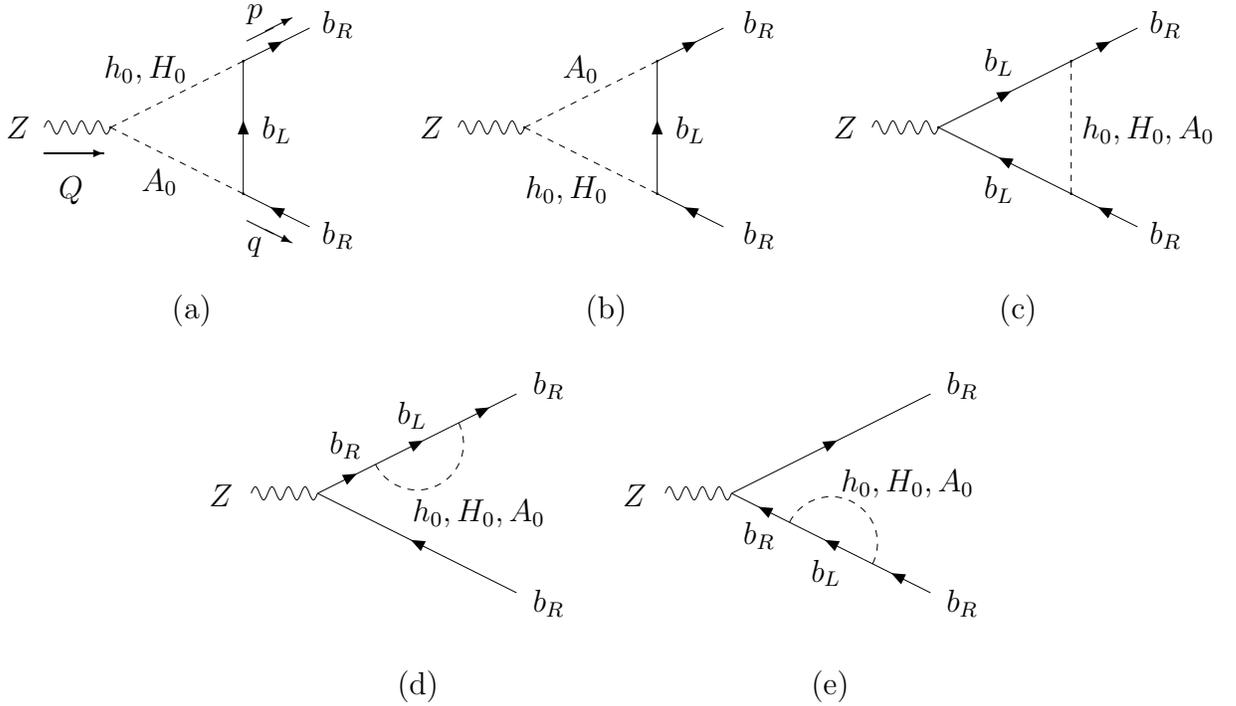}
\end{picture}
\caption{One--loop neutral Higgs corrections to
$Z\rightarrow b_R\bar{b}_R$.  Diagrams which correct 
$Z\rightarrow b_L\bar{b}_L$ can be obtained by the interchange
$b_L\leftrightarrow b_R$.} 
\label{neutralhiggs}
\end{center}
\end{figure}

%%%%%%%%%%%%%%%%%%%%%%%%%%%%%%%%%%%%%%%%%%%%%%%%%%%%%%%%%%%%%%%%%%%%%%%%%%%%%
\newpage

\begin{figure}[ht]
\begin{center}
\unitlength=1cm
\begin{picture}(17,11)(0,0)
\unitlength=1mm
%\zahyou{17}{11}
\put(13,77){\vector(1,0){8}}
\put(15,71){$Q$}
\put(40,92){\vector(2,1){6}}
\put(40,95){$p$}
\put(40,68){\vector(2,-1){6}}
\put(40,64){$q$}
\put(30,55){(a)}
\put(85,55){(b)}
\put(140,55){(c)}
\put(60,5){(d)}
\put(115,5){(e)}
\put(8,79){$Z$}
\put(63,79){$Z$}
\put(118,79){$Z$}
\put(42,79){$t_L$}
\put(97,79){$H^+$}
\put(152,79){$H^+$}
\put(35,31){$Z$}
\put(90,31){$Z$}
\put(50,94){$b_{R}$}
\put(50,65){$b_{R}$}
\put(106,94){$b_{R}$}
\put(106,65){$b_{R}$}
\put(160,94){$b_{R}$}
\put(160,65){$b_{R}$}
\put(78,45){$b_{R}$}
\put(78,16){$b_{R}$}
\put(133,45){$b_{R}$}
\put(133,16){$b_{R}$}
\put(26,86){$H^+$}
\put(26,71){$H^+$}
\put(82,88){$t_L$}
\put(82,72){$t_L$}
\put(132,86){$t_R$}
\put(132,74){$t_R$}
\put(142,90){$t_L$}
\put(142,70){$t_L$}
\put(56,46){$H^+$}
\put(48,37){$b_{R}$}
\put(63,34){$t_L$}
\put(103,26){$b_{R}$}
\put(111,14){$H^+$}
\put(119,27){$t_L$}
\epsfbox[0 470 480 780]{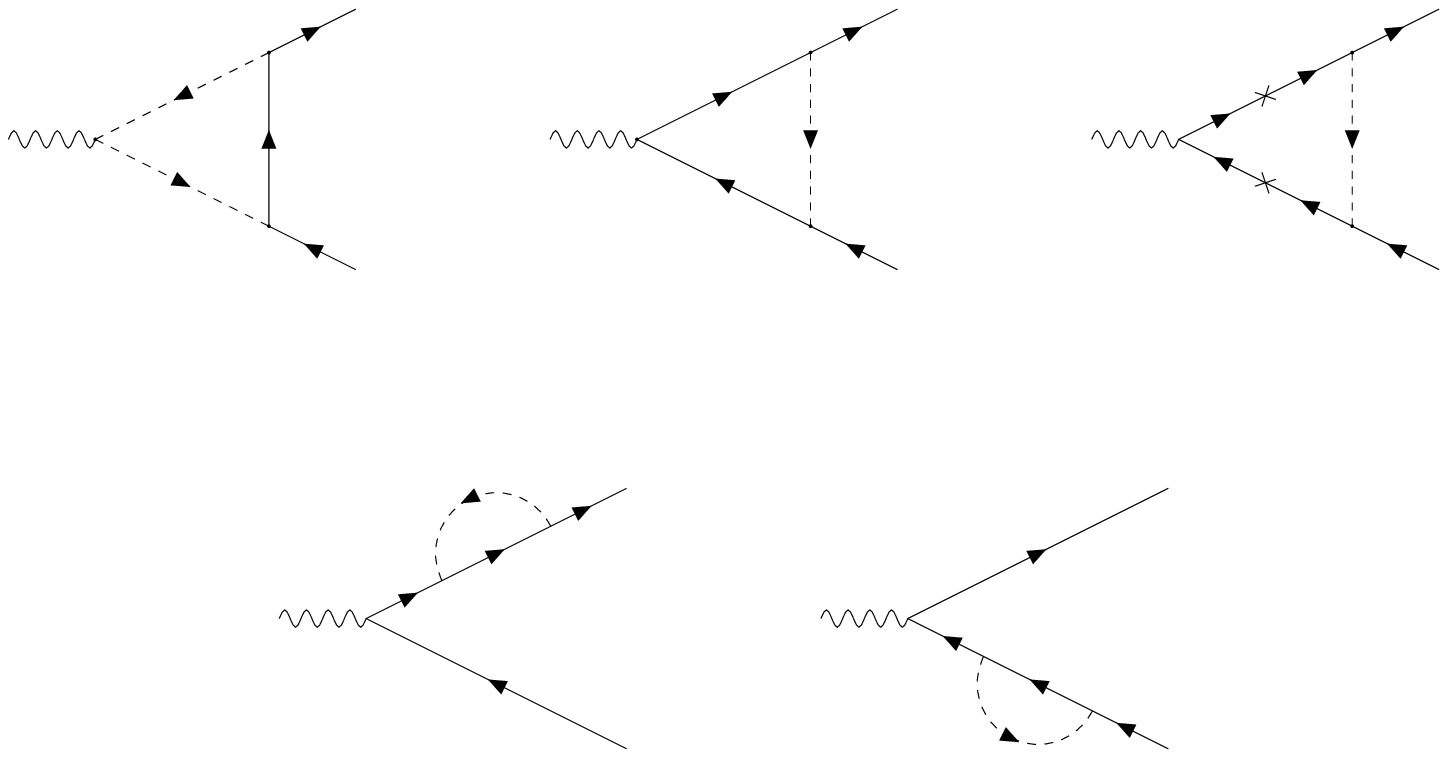}
\end{picture}
\caption{One--loop charged Higgs corrections to
$Z\rightarrow b_{R}\bar{b}_{R}$.   Diagrams which correct
$Z\rightarrow b_{L}\bar{b}_{L}$ can be obtained by the substitution
$b_R\rightarrow b_L$, $t_L\rightarrow t_R$.
} 
\label{chargedhiggs}
\end{center}
\end{figure}

%%%%%%%%%%%%%%%%%%%%%%%%%%%%%%%%%%%%%%%%%%%%%%%%%%%%%%%%%%%%%%%%%%%%%%%%%%%%%
\newpage

\begin{figure}[ht]
\begin{center}
\unitlength=1cm
\begin{picture}(13,10)(0,0)
\unitlength=1mm
%\zahyou{13}{10}
\put(40,55){(a)}
\put(96,55){(b)}
\put(40,5){(c)}
\put(96,5){(d)}
\put(17,78){$Z$}
\put(72,78){$Z$}
\put(17,29){$Z$}
\put(72,29){$Z$}
\put(50,78){$H^+$}
\put(46,27){$H^+$}
\put(102,31){$H^+$}
\put(105,78){$\tau_R$}
\put(40,40){$\tau_R$}
\put(96,19){$\tau_R$}
\put(30,35){$\nu_\tau$}
\put(86,24){$\nu_\tau$}
\put(59,92){$\nu_\tau$}
\put(114,92){$\nu_\tau$}
\put(59,64){$\nu_\tau$}
\put(114,64){$\nu_\tau$}
\put(59,43){$\nu_\tau$}
\put(114,43){$\nu_\tau$}
\put(59,15){$\nu_\tau$}
\put(114,15){$\nu_\tau$}
\put(36,87){$\tau_R$}
\put(36,70){$\tau_R$}
\put(92,87){$H^+$}
\put(92,68){$H^+$}
\epsfbox[0 0 360 290]{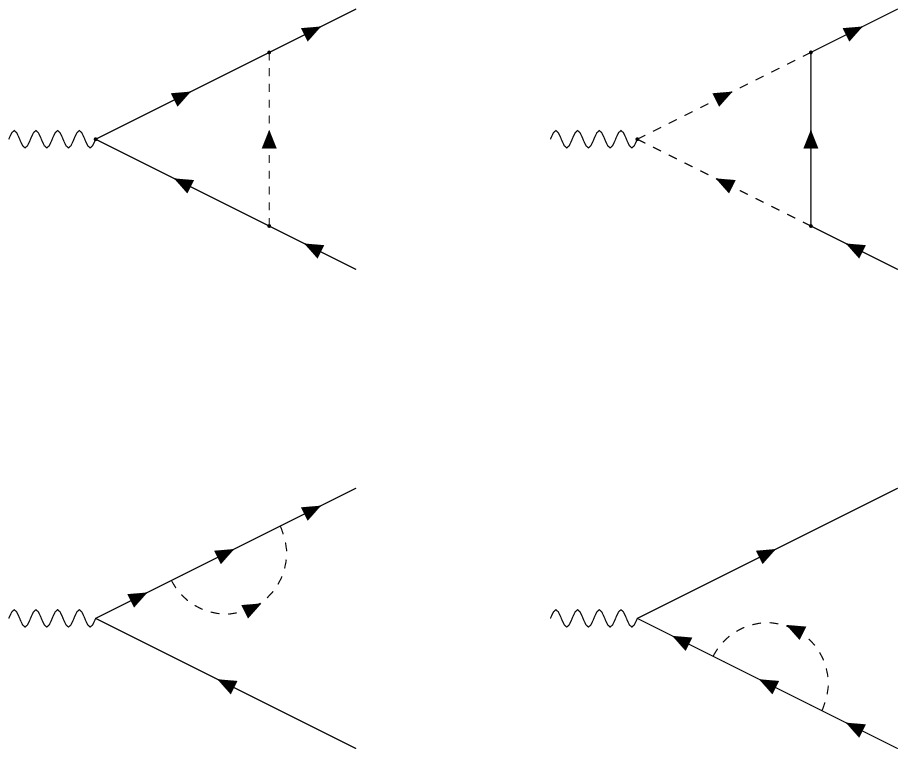}
\end{picture}
\caption{Charged Higgs corrections to $Z \rightarrow \nu_\tau 
{\overline\nu}_\tau$}
\label{neutrinodiagrams}
\end{center}
\end{figure}

%%%%%%%%%%%%%%%%%%%%%%%%%%%%%%%%%%%%%%%%%%%%%%%%%%%%%%%%%%%%%%%%%%%%%%%%%%%%%
\newpage

\begin{figure}
\begin{center}
\begin{picture}(12,10)(0,0)
\unitlength=1mm
%\zahyou{12}{10}
\put(92,51){$A_{\rm FB}(b)$}
\put(82,26){$R_\tau$}
\put(56,32){$R_b$}
\put(51,25){$R_\mu$}
\put(29,33){$R_e$}
\put(35,84){$A_b$}
\put(64,2){$\delta h_{\tau_L}^{N}$}
\put(0,56){$\delta h_{b_R}^{C}$}
\epsfbox[-30 -20 290 270]{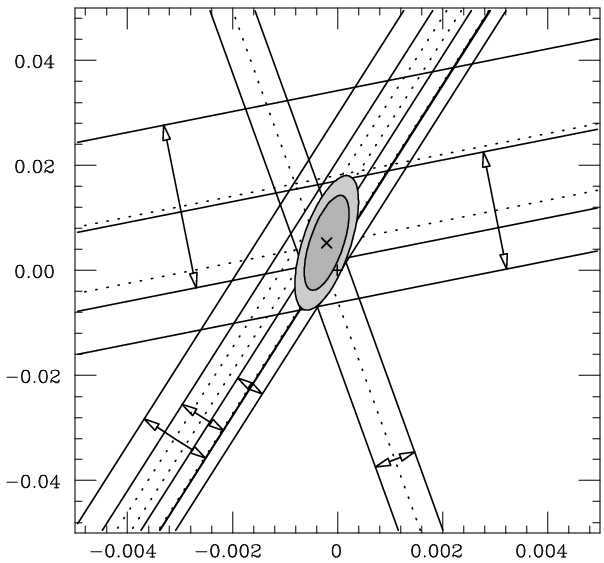}
\end{picture}
\caption{The $1\sigma$ constraints on $\delta h_{\tau_L}^{N}$ and
$\delta h_{b_R}^{C}$ from various observables in the
$\delta s^2 = \delta\alpha_s = 0$ plane.
The shaded contours represent the 68\% and 90\% confidence
limits.}
\label{Contour1}
\end{center}
\end{figure}

\begin{figure}
\begin{center}
\begin{picture}(12,10)(0,0)
\unitlength=1mm
%\zahyou{12}{10}
\put(77,87){$A_\tau({\rm LEP})$}
\put(46,87){$A_{\rm LR}$}
\put(71,24){$A_{\rm FB}(b)$}
\put(36,66){$R_\mu$}
\put(89,28){$R_e$}
\put(102,47){$R_b$}
\put(91,65){$R_\tau$}
\put(65,2){$\delta s^2$}
\put(0,56){$\delta h_{\tau_L}^{N}$}
\epsfbox[-30 -20 290 270]{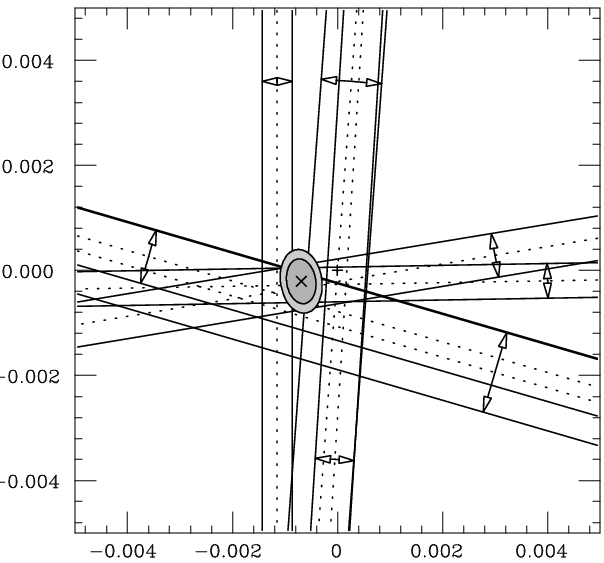}
\end{picture}
\caption{The $1\sigma$ constraints on $\delta h_{\tau_L}^{N}$ and
$\delta s^2$ from various observables in the
$\delta h_{b_R}^{C} = \delta\alpha_s = 0$ plane.
The shaded contours represent the 68\% and 90\% confidence
limits.}
\label{Contour2}
\end{center}
\end{figure}

%%%%%%%%%%%%%%%%%%%%%%%%%%%%%%%%%%%%%%%%%%%%%%%%%%%%%%%%%%%%%%%%%%%%%%%%%%%%%
\newpage

\begin{figure}
\begin{center}
\begin{picture}(12,10)(0,0)
\unitlength=1mm
%\zahyou{12}{10}
\put(37,78){$A_b$}
\put(38,42){$R_\mu$}
\put(39,24){$R_\tau$}
\put(62,20){$A_{\rm LR}$}
\put(85,22){$A_{\rm FB}(b)$}
\put(102,50){$R_b$}
\put(84,94){$R_e$}
\put(65,2){$\delta s^2$}
\put(0,56){$\delta h_{b_R}^{C}$}
\epsfbox[-30 -20 290 270]{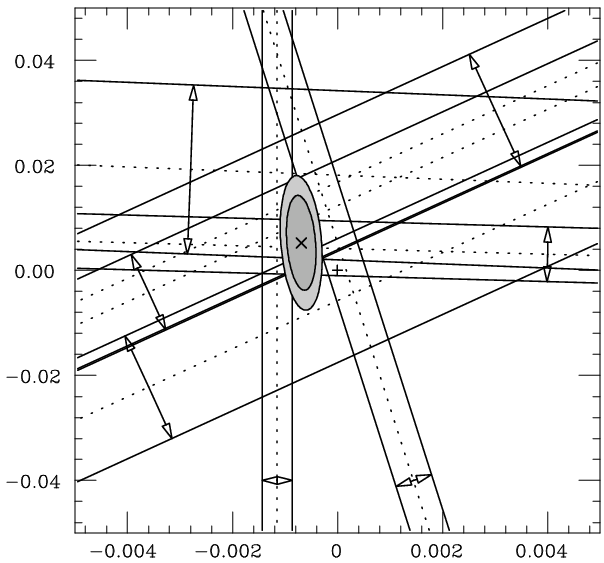}
\end{picture}
\caption{The $1\sigma$ constraints on $\delta h_{b_R}^{C}$ and
$\delta s^2$ from various observables in the
$\delta h_{\tau_L}^{N} = \delta\alpha_s = 0$ plane.
The shaded contours represent the 68\% and 90\% confidence
limits.}
\label{Contour3}
\end{center}
\end{figure}

%%%%%%%%%%%%%%%%%%%%%%%%%%%%%%%%%%%%%%%%%%%%%%%%%%%%%%%%%%%%%%%%%%%%%%%%%%%%%
\newpage

\begin{figure}
\begin{center}
\epsfbox[0 0 340 260]{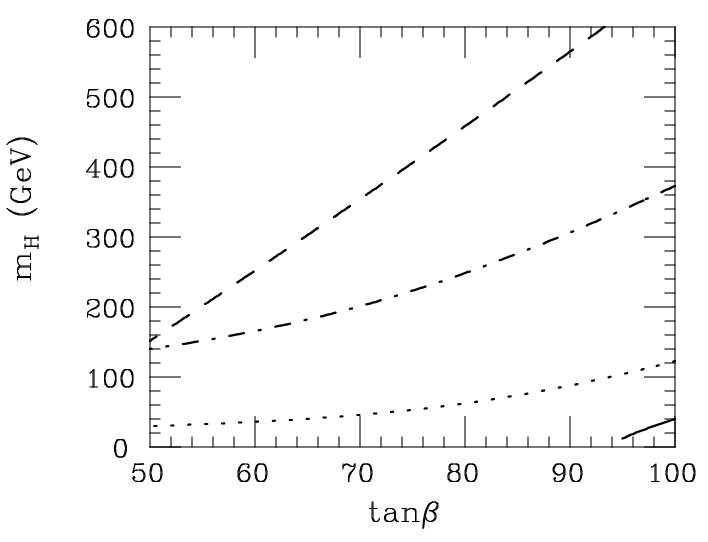}
\caption{Lower bounds on the charged Higgs mass vs. $\tan\beta$.  
The dot--dashed and dotted lines correspond to the Bayesian
68 \% and 95 \% confidence levels, respectively.  
The dashed and solid lines correspond to the 
classical 68 \% and 95 \% confidence levels, respectively.}
\label{chargedhiggsmass}
\end{center}
%\end{figure}

%%%%%%%%%%%%%%%%%%%%%%%%%%%%%%%%%%%%%%%%%%%%%%%%%%%%%%%%%%%%%%%%%%%%%%%%%%%%%

%\begin{figure}
\begin{center}
\epsfbox[0 0 340 260]{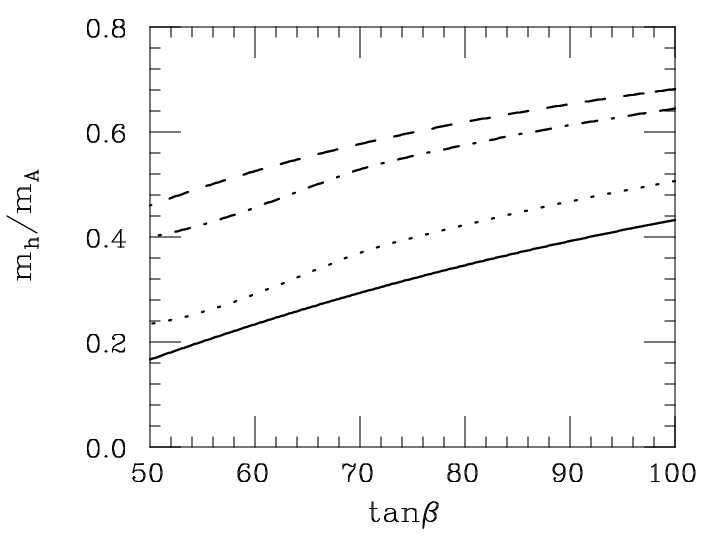}
\caption{Lower bounds on the scalar--pseudoscalar mass ratio vs. $\tan\beta$.  
The dot--dashed and dotted lines correspond to the Bayesian
68 \% and 95 \% confidence levels, respectively.  
The dashed and solid lines correspond to the 
classical 68 \% and 95 \% confidence levels, respectively.}
\label{neutralhiggsmass}
\end{center}
\end{figure}

%%%%%%%%%%%%%%%%%%%%%%%%%%%%%%%%%%%%%%%%%%%%%%%%%%%%%%%%%%%%%%%%%%%%%%%%%%%%%
%\narrowtext

%%%%%%%%%%%%%%%%%%%%%%%%%%%%%%%%%%%%%%%%%%%%%%%%%%%%%%%%%%%%%%%%%%%%%%%%%%%%%
\end{document}